\documentclass{aastex}
\usepackage{amsmath}
\usepackage{graphicx}
\usepackage{natbib}

\shorttitle{Separated Fringe Packet Binaries II}
\shortauthors{Farrington et al.}

\begin{document}

\title{Separated Fringe Packet Observations with the CHARA Array II: $\omega$~Andromeda, HD~178911, and $\xi$~Cephei.}


\author{C.D. Farrington, T.A. ten Brummelaar}
\affil{The CHARA Array, Mt. Wilson Observatory, Mt. Wilson, CA 91023}
\email{farrington@chara-array.org, theo@chara-array.org}

\author{B.D Mason, W.I Hartkopf}
\affil{US Naval Observatory, 3450 Massachusetts Avenue NW, Washington, DC 20392-5420}
\email{bdm@usno.navy.mil, wih@usno.navy.mil}

\author{D. Mourard}
\affil{Universit\'{e} de Nice Sophia Antipolis, CNRS, Laboratoire J.L.
Lagrange, Observatoire de la C\^{o}te d'Azur - BP4209, 06304 Nice cedex, 
France}
\email{denis.mourard@oca.eu}

\author{E. Moravveji}
\affil{Instituut voor Sterrenkunde, KU Leuven, Celestijnenlaan 200D, 3001, 
Leuven, Belgium}
\email{Ehsan.Moravveji@ster.kuleuven.be}

\author{H.A. McAlister}
\affil{Center for High Angular Resolution Astronomy, Georgia State University, P.O. Box 3969, Atlanta, GA 30302-3969}
\email{hal@chara.gsu.edu}

\author{N.H. Turner, L. Sturmann, J. Sturmann}
\affil{The CHARA Array, Mt. Wilson Observatory, Mt. Wilson, CA 91023}
\email{nils@chara-array.org, sturmann@chara-array.org, judit@chara-array.org}

\begin{abstract}{}

 When observed with optical long-baseline interferometers (OLBI), components of a binary star which are sufficiently separated produce their own interferometric fringe packets; these are referred to as Separated Fringe Packet (SFP) binaries. These SFP binaries can overlap in angular separation with the regime of systems resolvable by speckle interferometry at single, large-aperture telescopes and can provide additional measurements for preliminary orbits lacking good phase coverage, help constrain elements of already established orbits, and locate new binaries in the undersampled regime between the bounds of spectroscopic surveys and speckle interferometry.  In this process, a visibility calibration star is not needed, and the separated fringe packets can provide an accurate vector separation.  In this paper, we apply the SFP approach to $\omega$~Andromeda, HD~178911, and $\xi$~Cephei with the CLIMB three-beam combiner at the CHARA Array. For these systems we determine component masses and parallax of 0.963$\pm$0.049 $M_{\odot}$ and 0.860$\pm$0.051 $M_{\odot}$ and 39.54$\pm$1.85 milliarcseconds (mas) for $\omega$~Andromeda, for HD~178911 of 0.802$\pm$0.055 $M_{\odot}$ and 0.622$\pm$0.053 $M_{\odot}$ with  28.26$\pm$1.70 mas, and masses of 1.045$\pm$0.031 $M_{\odot}$ and 0.408$\pm$0.066 $M_{\odot}$ and 38.10$\pm$2.81 mas for $\xi$~Cephei. 
\end{abstract}


\vskip 0.5pt

\section{Introduction}

Long-baseline interferometric telescope arrays are well-suited for observing binaries with angular separations in the sub-millarcsecond regime using the traditional interferometric visibility method [for examples, see \citet{ARMSTRONG}, \citet{BODEN}, \citet{HUMMEL}, and \citet{RAG2009}].  Another approach [\citet{DYCK}, \citet{LANE}, \citet{BAG}, \citet{CHARA96}] applies to stellar systems where the components of a binary are sufficiently far apart in projected angular separation that their fringe packets do not overlap and the visibility fitting approach is not relevant. This paper follows \citet{FAR2010} (hereafter referred to as Paper I) presenting the results from a program of separated fringe packet (SFP) observations of spectroscopic and visual binary star systems made with the CHARA Array at Mount Wilson Observatory \citep{CHARA2}.  Paper I contained the systems $\chi$~Draconis, HD~184467, and HD~198084, and presented new observations, orbits, and masses for each system, and a variant of this technique is presented for triple systems in \citet{OBR2011}.  As part of this ongoing effort, we present here 150 new vector measurements of $\omega$~Andromeda, HD~178911, and $\xi$~Cephei that are combined into 60 positional observations of the components of these systems. With this second paper, we have refined the process of data collection and reduction to incorporate the increased capacity and efficiency of the CLIMB (CLassic Interferometry with Multiple Baselines) beam combiner \citep{JAI}.

\section{Observational Overview}

Data were routinely taken on the CHARA Array's three largest baselines (S1-E1-W1) and other intermediate baselines when the preferred telescopes were assigned to other simultaneous observing experiments. A list of observations for $\omega$~Andromeda, HD~178911, and $\xi$~Cephei taken with the CHARA Array, along with baselines used, is given in Table \ref{tab_sfpobs}. This table contains the acquired 1-D measurements in columns $3-6$ and the 2-D positional calculation obtained from the observations in columns $7-9$. Each 1-D measurement consist of averaged time, length of baseline, and position angle of the projected baseline at the midpoint of the five minute recording sequence and a separation between the peaks of the two average fringe envelopes that have been summed over the course of the data file [see \citet{FAR2010}]. The 2-D columns represent the combination of all the 1-D data for a given set of observations through the program described in the Data Reduction section below. The exceptions to the above descriptions are those data labeled "VEGA'' in the table.  These measurements do not use the SFP method but visibility modulation typical for interferometers, and thus do not consist of 1-D vector measurements. Full details of the VEGA instrument can be found in \citet{VEGA}.

Before 2009, observations for the SFP program were taken as described in Paper I with the CHARA Classic two-beam combiner as described by \citep{CHARA2}. All observations after 2009 were taken with the CHARA CLIMB IR pupil-plane three-beam combiner \citep{JAI} also through the $K'$ filter.  The timespan between observing sessions ranged from as little as a week to more than a year.  Orbits for these systems were determined with combined spectroscopic/interferometric solutions as described in \citet{Toko1992,Toko1993} with all available CHARA, published speckle interfrometry data \citep{FOURTH}, and spectroscopic orbits as described below.

\subsection{Characterizing Separated Fringe Packets}
 The theory, history, errors, and method of utilizing SFP interferometry are discussed in detail in Paper I. Several important changes have been implemented since the publication of that paper that have increased the accuracy, quality, and speed of the data acquisition with the CHARA Array.  In 2009, the new CLIMB three-beam combiner \citep{INSTRUMENT} was built alongside of the previously used CLASSIC two-beam combiner. While primarily built for multiple simultaneous baseline observations to determine closure phase \citep{JAI}, the SFP project found an alternative use for the combiner, as the primary mode for CLIMB used two dither mirrors working simultaneously at different frequencies and movement parameters narrowed the delay-space being sampled at any given time.  If used in its primary mode, this would decrease the 1-D sky coverage of two of the baselines by 25\% for the second pair of baselines that include a dither mirror, and 50\% for the final pair which is considered the ``cross fringe."  In order to retain the largest possible sky coverage, a two-beam mode was added that used the same frequencies and largest possible delay-space search for all three baselines, but only recorded one baseline at a time.  With this mode on CLIMB, the amount of time needed to observe one object on all three baselines took less than a quarter of the time required by the method described in Paper I.          

\subsection{Data Reduction}
 Most of the data reduction was done with the same method and software as described in \citet{FAR2010} with the exception of the final stage, the determination of the 2-D location of the companion. 

\subsubsection{Calculation Method for Astrometry from SFP Data}

Each observation of a binary star produces a linear separation of the system on the sky, whose direction is determined by the projection angle of the baseline on the sky and whose distance is determined by the separation of the two fringe packets divided by the projected baseline length \citep{FAR2010}. Thus, if we place the primary, as defined by the star that produces the largest fringe packet\footnote{Note that the star that produces the largest fringe packet is not necessarily the brightest star as the brightest star may be more resolved at the current baseline than the fainter star and its fringe packet is suppressed.}, at the origin, each observation will produce a line, which for observation $i$ we write as
\begin{equation}
y = m_i x + c_i.
\end{equation}
For any single observation, the position of the secondary 
$(x_{\rm s}, y_{\rm s})$ can 
lie anywhere on this line, but for more than one observation the position of the secondary is more restricted. Ideally all of the lines will intersect at the position of the secondary, but of course the presence of noise makes this extremely unlikely. We therefore use the equivalent of a $\chi^2$ minimization.

\begin{figure}[!t]
  \centering \rotatebox{-90}{\includegraphics[width=0.7\textwidth]
  {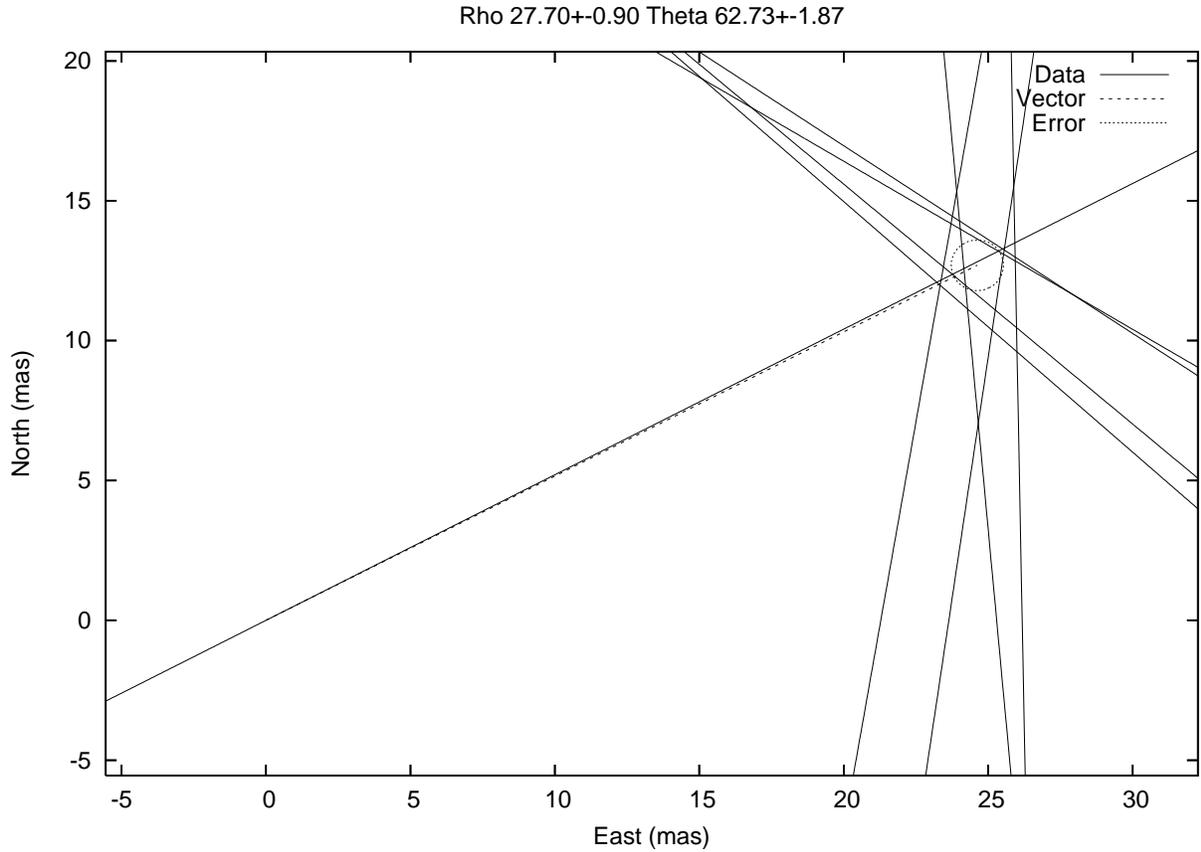}}\\
\vspace{0.2in}
  \caption[Example Vector Separation Plot for $\omega$~Andromeda, 2012.5615]
  {\footnotesize{Example Vector Separation Plot for $\omega$~Andromeda, 2012.5615. An example output from the SFPAstrom program that solves for the location of the companion from multiple 1-D vector measurements. The dashed line is the vector from the origin to the estimated location of the companion.  Each solid line is one 1-D vector measurement, and the dashed circle is the error ellipse for the best estimate for the position of the secondary and the size of the ellipse represents the error of the secondary position. }}
  \label{sfpastrom_fig}
\end{figure}

We write the distance of the 
secondary from the line defined by observation $i$ as
\begin{equation}
\Delta_i = \sqrt{(x_{\rm s} - x_p)^2 + (y_{\rm s} - y_p)^2},
\end{equation}
where $(x_p, y_p)$ is the point on the line defining the perpendicular distance between $(x_{\rm s}, y_{\rm s})$ and the line given by
\begin{equation}
x_p = \frac{x_{\rm s} + m_i y_{\rm s} - m_i c_i}{1 + m_i^2}
\end{equation}
and 
\begin{equation}
y_p = m_i x_p + c_i.
\end{equation}
We then say that the $\chi^2$ of any secondary position is given by
\begin{equation}
\chi^2 = \sum_{i=1}^{N_{\rm obs}} \frac{\Delta_i^2}{\sigma_i^2}
\end{equation}
where $\sigma_i^2$ is the variance of the linear separation of
observation $i$. Following standard $\chi^2$ analysis, we say that the best estimate for the position of the secondary is given by the values of 
$(x_{\rm s}, y_{\rm s})$ that minimize $\chi^2$. 
Since this is not a true $\chi^2$ measurment the error can not be estimated in
the normal way. Instead we use the standard deviation of the perpendicular 
distances, $\Delta_i$. A program was written in C by T.A. ten Brummelaar that 
does the above calculations called ``SFPAstrom'' and a sample output of the resulting fits is displayed in Figure \ref{sfpastrom_fig}. 

\subsection{Effects of Misalignment}
In Paper I, the most prevalent possible sources of error were discussed and all but the piston error were of such a small magnitude that they could essentially be dismissed.  It is worth quantifying the potential error in separation of two fringe packets brought about by the misalignment of the optical path from the beam combiner out to the telescope on one arm of the interferometer.  

\begin{figure}[!t]
  \centering  \includegraphics[width=0.5\textwidth]
  {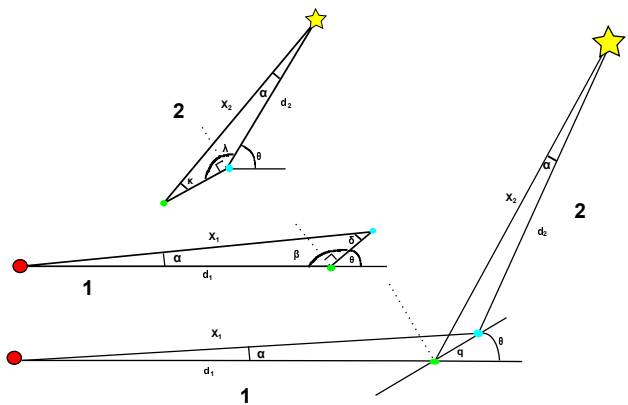}\\
\vspace{0.2in}
  \caption[Beam Misalignment Diagram]
  {\footnotesize{Simple diagram of the components comprising beam path from the beam combiner to the observed system. The distances $\chi_{1}$ and $\chi_{2}$ are the misaligned paths when $d_{1}$ and $d_{2}$ are the optical axis when the alignment is done correctly, $\alpha$ is the angle subtended by the path difference between the beam combiner and the telescope, and $\theta$ is the angle of the telescope with $0$ at zenith and $90$ at the horizon.  The difference between the $\chi$ and $d$ paths is calculated in terms of $\alpha$ and $\theta$.}}
  \label{misdelay}
\end{figure}

\begin{figure}[!t]
  \centering  \includegraphics[width=0.5\textwidth]
  {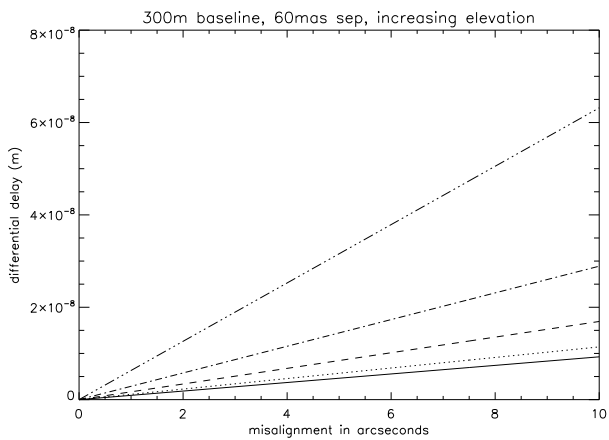}\\
\vspace{0.2in}
  \caption[Differential Delay vs. Beam Misalignment]
  {\footnotesize{Calculated differential delay based on telescope angle and realistic beam misalignment. The lines correspond to angles from near the horizon to near zenith (solid line near zenith, each line after decreases the zenith angle by 15 degrees, ending at 30 degrees above the horizon) and represent the difference between the single and double star cases described in Equation \ref{diff} compared to the ideal path case of approximately 1$\mu$m.}}
  \label{misdelay2}
\end{figure}

Starting with the configuration in Figure \ref{misdelay}, we can calculate the error in path for a single star for a typical misalignment that could occur due to coud\'{e} variation in azimuth of approximately 5mm or about 10 arcseconds over the longest baseline.  We want to determine $\chi_{1}$ and $\chi_{2}$ in terms of the nominal distances ($d_{1}$ and $d_{2}$), the angle of the telescope, $\theta$, and the misalignment angle, $\alpha$.  From simple geometric identities, it can be shown that:
\begin{equation}
\chi_1 = \frac{d_1 \sin(\frac{\alpha}{2} - \frac{\theta}{2})}{\sin(\frac{3 \alpha}{2} - \frac{\theta}{2})}
\end{equation}
and
\begin{equation}
\chi_2 = \frac{d_2 \sin(\frac{3 \alpha}{2} + \frac{\theta}{2})}{\sin(\frac{\alpha}{2} + \frac{\theta}{2})}
\end{equation}
For the pertinent case where we are observing a binary system that would produce two fringes as described in Paper I, we show that the path difference for the individual components for a relatively wide realistic case:
\begin{equation}
\chi_{1}(\theta) + \chi_{2}(\theta) - \chi_{1}(\theta + \Delta) -\chi_{2}(\theta + \Delta) \ll \vec{\rho}_{\rm \mu m }
\label{diff}
\end{equation}
where $\Delta$ is the on-sky separation of the two fringes in milliarcseconds
and rearranging Equation 5 from Paper I gives:
\begin{equation}
\vec{\rho}_{\rm \mu m } = \frac{\vec{\rho}_{\rm mas} B(m)}{206.265} ~~.
\label{sep}
\end{equation} 
The solutions for Equation \ref{diff} are given in Figure \ref{misdelay2} with increasing $\theta$ and from 0-10 arcsecond misalignment for a baseline of 300~m and projected binary star separation of 60~mas, show a maximum differential delay that is 3 orders of magnitude smaller than the separation between the two fringes (for this example, the separation of the fringes in microns is approximately 87$\mu$m, and the error due to the largest misalignment approaches 0.08$\mu$m), and thus far smaller than atmospheric piston, the most dominant source of positional error. 

\section{Results}
\subsection{$\omega$~Andromeda}
\begin{figure}[!t]
  \centering  \includegraphics[width=0.7\textwidth]
  {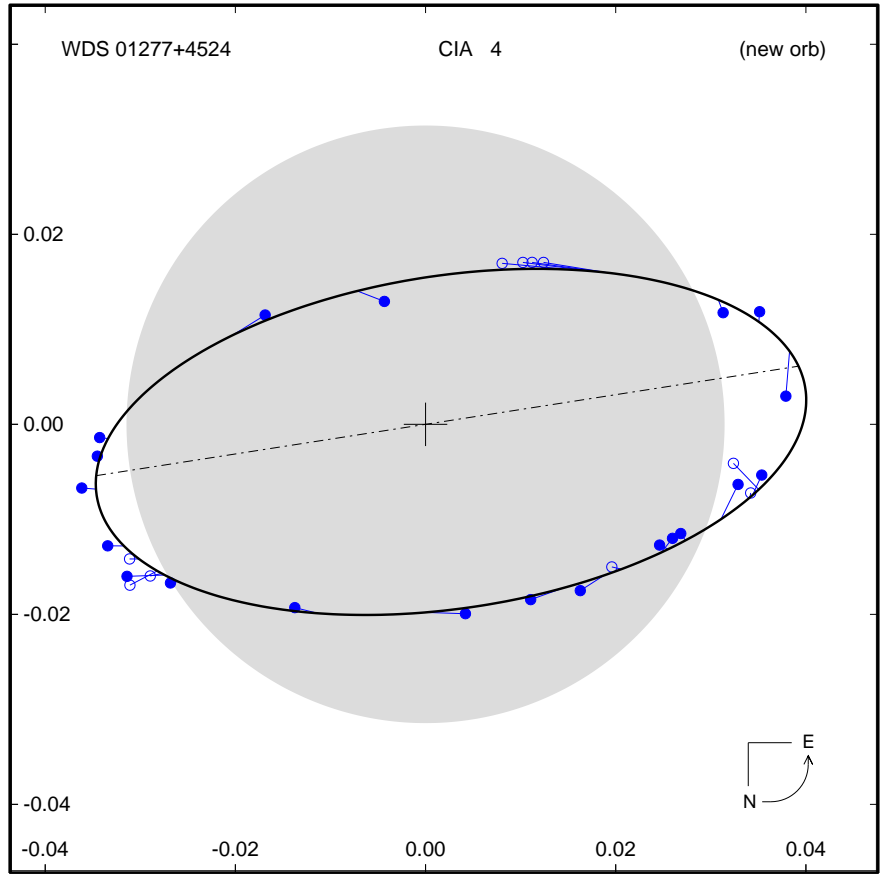}\\
\vspace{0.2in}
  \caption[Orbit Plot for $\omega$~Andromeda]
  {\footnotesize{Orbit Plot for $\omega$~Andromeda. Combined visual-spectroscopic solution from this paper (solid line) using all available data which is all from the CHARA Array SFP Program. The shaded circle represents the resolution limit of a speckle interferometry camera on a 4-m telescope and is shown to aid in scaling. The dot-dash line indicates the line of nodes. The VEGA beam combiner measures are shown as open circles. The CHARA Array SFP measures are indicated with filled circles. All measurements are connected to their predicted positions on the orbit by ``O$-$C'' lines. The direction of motion is indicated on the north-east orientation in the lower right of the plot. The scales at left and bottom are in arcseconds.}}
  \label{8799fig}
\end{figure}

$\omega$~Andromeda = HR~417, HD~8799, spectral types are suggested to be F3V+F5V \citep{ABT2,COWLEY}. The listed B component is faint (12th magnitude at 2$''$) and may be optical \citet{BURN1,BURN2}.  The system also contains a second pair 2$'$ distant, separated by 5$''$ with a combined magnitude of 10, designated as components CD, which are optical. No previous astrometric or interferometric observations of the system have been published.  All astrometric data taken for this system was obtained on the CHARA Array using CLIMB and the VEGA visible beam combiner \citep{VEGA}. VEGA data are not processed through the SFP principle but they used the classical principle of visibility modulation as a function of time, baseline, as in \citet{PAN90}. The spectroscopic orbit used in the combined solution presented here is from \citet{GRIFFIN}.  A simultaneous solution utilizing all the radial velocity and visual data was carried out with an interactive program developed by \citet{Toko1992,Toko1993} that computes all 10 orbital elements. This technique employs the method of least squares to yield elements satisfying both radial velocity and astrometric measurements as described in \citet{MCA1995}. The orbital elements from the combined solution are listed in Table \ref{tab_8799}, along with the orbital $\chi_{\nu}^2$, masses, and orbital parallax calculated from the solution, and Figure \ref{8799fig} shows the relative orbit.  The orbital parallax of 39.54$\pm$1.85 milliarcseconds (mas) is different from that of Hipparcos (34.94$\pm$0.31 mas; \citet{HIP2}), probably due to the pair being unresolved and the parallax being biased with the binary separation.  The calculated masses are 0.963$\pm$0.049 $M_{\odot}$ and 0.860$\pm$0.051 $M_{\odot}$ for the components and an orbital grade of $1$ determined by criteria of the Sixth Orbital Catalog \citep{ORB6}.

\subsection{HD~178911}
\begin{figure}[!t]
  \centering  \includegraphics[width=0.7\textwidth]
  {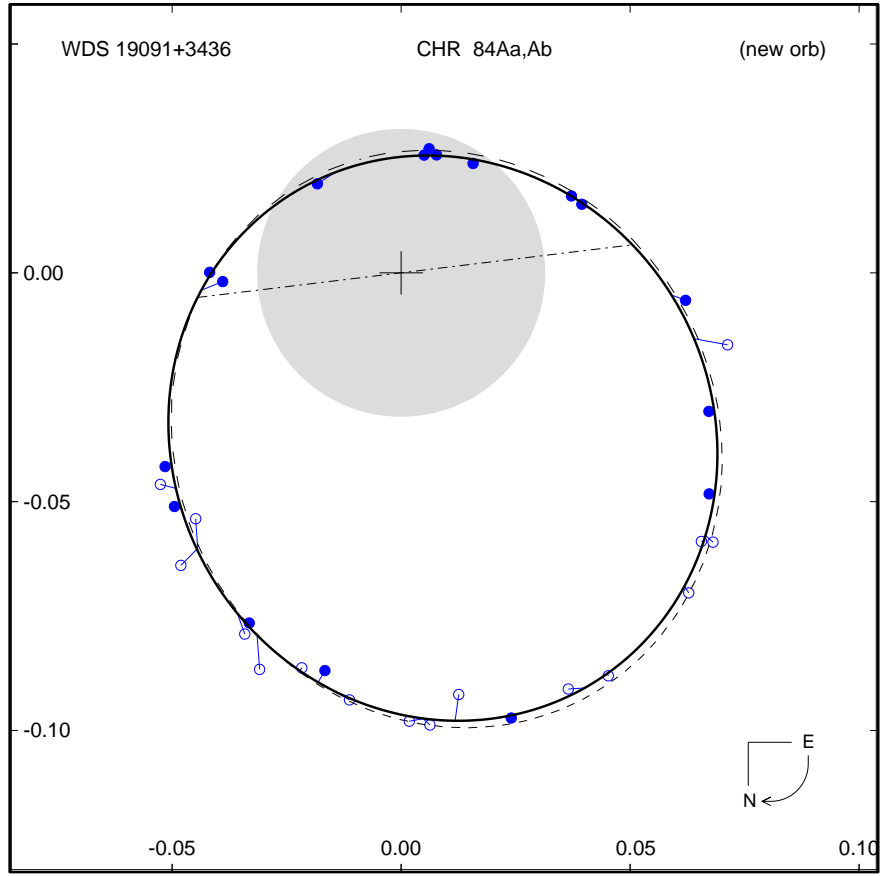}\\
\vspace{0.2in}
  \caption[Orbit Plot for HD~178911]
  {\footnotesize{Orbit Plot for HD~178911. Combined visual-spectroscopic solution from this paper (solid line) using all available data which is consistent with the previous orbit of \citet{Toko2000} (dashed line). The shaded circle represents the resolution limit of a speckle interferometry camera on a 4-m telescope. The CHARA Array SFP measures are indicated with filled circles. Speckle interferometry measurements are indicated as open circles. Other symbols as Figure \ref{8799fig}.}}
  \label{178911fig}
\end{figure}

HD~178911 = HR 7272, CHR 84Aa,Ab, spectral types are G1V+K1V.  Measured diameter is 0.114 mas \citep{RIBAS}. The AB pair (16$''$, $\Delta$m=1.1) is known as STF2747 and shares a common proper motion. The B component of the wide pair is an extrasolar planet host star \citep{WITT2009}. The much wider AC pair is known as WAL 105 (96$''$, $\Delta$m=4.6) and is optical. The close Aa,Ab pair was discovered by the CHARA speckle interferometry program in 1985 \citep{MCA1987}. Summary information for components can be found in the Washington Double Star (WDS) Catalog \citep{WDS}. The entry from the Sixth Visual Orbit Catalog [ORB6; \citet{ORB6}] is from \citet{HRT2000}, and the spectroscopic orbit and first combined solution are from \citet{Toko2000}. While the previous orbit included only six visual measurements, our solution is quite similar with reduced errors while including 17 measurements from the CHARA Array, and 10 other subsequent speckle interferometric data points. This five-fold increase in the number of measures of relative astrometry has a significant impact on the mass and other determinations due to much lower errors. The orbit, presented in Table \ref{tab_7272}, was computed using the same combined solution technique of \citep{Toko1992,Toko1993} listed above deriving all ten orbital parameters as well as orbital $\chi_{\nu}^2$, component masses, and orbital parallax.  Figure \ref{178911fig} plots the previous and current orbital solutions with all measurements previous to this effort.  The orbital parallax of 28.26$\pm$1.70 milliarcseconds (mas) is different from that of Hipparcos (19.11$\pm$2.35 mas; \citep{HIP2}), probably due to the pair being unresolved and the parallax being affected by the binary separation. Using the objective orbit grading scheme described in ORB6 a grade of 1, definitive, has been determined for this pair. As with all other orbits in ORB6, this is based only on the orbital elements and the resolved measures and, therefore, does not take into account the spectroscopic solution which significantly improves the quality. The calculated masses of 0.802$\pm$0.055 $M_{\odot}$ and 0.622$\pm$0.053 $M_{\odot}$ for the Aa and Ab components, while lower, are within the error margin of the previous solution.  

\subsection{$\xi$~Cephei A}
$\xi$~Cephei A = HR~8417, HD~209790, MCA 69Aa,Ab. The AB (5-8$''$, $\Delta$m=2.0) and AC (110$''$, $\Delta$m=8.2) pairs are both known as STF2863. B shares a commmon proper motion with A. The C component has only been measured a few times since its discovery in 1925 \citep{OPIK} and its status, whether optical or physical, is unknown. \citet{EGGEN1,EGGEN2} has determined the system to be a member of the IC 2391 supercluster. Summary information for the system can be found in the Washington Double Star (WDS) Catalog \citep{WDS}. \citet{HYNEK} included the close pair in a list of composite spectrum binaries, and \citet{ABT} suggested that it is long-period spectroscopic binary. \citet{VS1976} confirmed Abt's suspicion, finding the system to be double-lined with an orbital period of 811 days. From an analysis of colors, Vickers \& Scarfe assigned spectral types of A7 for Aa and F5 for Ab, suspecting on the basis of strong lines of strontium and ionized iron that the secondary is a subgiant. The fit to the colors leads to a $\Delta$V = 0.3 magnitudes, a value significantly smaller than that expected for a pair of A7 and F5 dwarfs, lending further support to the evolved nature of the secondary. Vickers \& Scarfe also measured the radial velocity of the B component and found it indistinguishable from the $\gamma$-velocity of the Aa,Ab system, confirming the common proper motion physicality. The system Aa,Ab was subsequently resolved by speckle interferometry \citep{MCA1977} and in \citet{MCA1980} the first relative orbit was derived from ten speckle observations and compared with the spectroscopic orbit of \citep{VS1976}.  

\begin{figure}[!t]
  \centering  \includegraphics[width=0.7\textwidth]
  {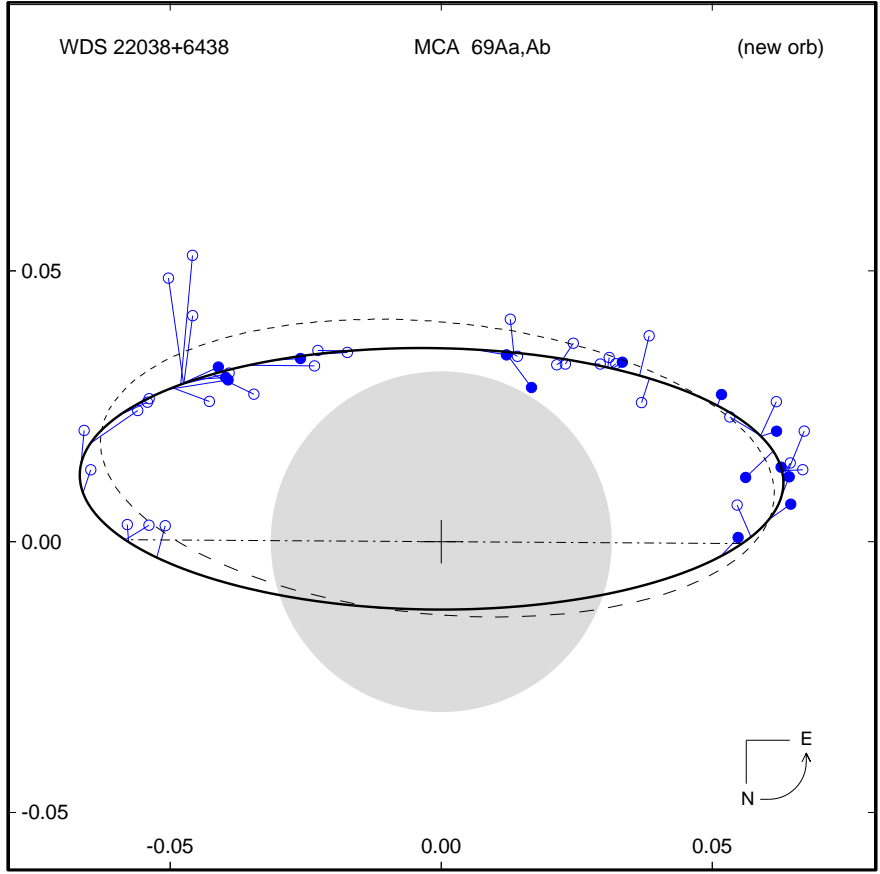}\\
\vspace{0.2in}
  \caption[Orbit Plot for $\xi$~Cephei]
  {\footnotesize{Orbit Plot for $\xi$~Cephei. The figure shows the 
relative visual orbit of the system; the x and y scales are in arcseconds. 
The solid curve represents the orbit determined in this paper with the dashed curve denoting the orbit of \citet{POURBAIX}. Other symbols as Figure \ref{178911fig}. }}
  \label{209790fig}
\end{figure}

The passage of time has quadrupled the number of interferometric measurements, most recently in the separated fringe packet campaign with the CHARA Array, and more importantly, this has increased the phase coverage from 1.3 to 16.4 orbital revolutions. All published observations of the pair are listed in the Fourth Interferometric Catalog \citep{FOURTH} including the recent measures by speckle interferometry by \citet{HORCH2008, HORCH2010}. The orbit, as presented in Table \ref{tab_209790} was computed using the same combined solution technique of \citep{Toko1992,Toko1993} listed above and plotted in Figure \ref{209790fig}. As above, the orbital parallax of 38.10$\pm$2.81 mas is different from that of Hipparcos (33.79$\pm$1.06 mas; \citep{HIP2}), probably due to the pair being unresolved and the parallax again being biased by the binary separation. Using the objective orbit grading scheme described in ORB6 a grade of 2, good, has been determined for this pair. As above this is based only on the orbital elements and the resolved measures and does not take into account the spectroscopic solution which significantly improves the quality. The masses of 1.045$\pm$0.031 $M_{\odot}$ and 0.408$\pm$0.066 $M_{\odot}$ for the components are of the same order as the previous solutions but are significantly different from what should be expected from a system with spectral types listed above.

\section{Conclusion}
As it was suggested in the first paper of this series, the inclusion of the CLIMB beam combiner did significantly increase the accuracy and alacrity of data acquisition for the SFP binary program.  The three systems observed in this paper are just the first of many that are available to this technique and the ongoing effort continues to add new spectroscopic binaries that are within the available observation range for orbit determination.  It should be noted that for the three systems discussed herein, and $\chi$ Draconis from Paper I of this series, the combined orbital solutions provide masses that do not mesh well with the predicted masses assigned from spectral typing. We present these orbits as they are computed, without prejudice to previously quoted spectral types, as the spectral typing and luminosity class determination are beyond the scope of the current investigation.  Additionally, five of the six objects from both this discussion and Paper I show significant differences between the orbital parallax calculated here and the Hipparcos parallax measurements due to the binarity unresolved at that time.

\acknowledgments

The CHARA Array, operated by Georgia State University, was built with funding provided by the National Science Foundation, Georgia State University, the W. M. Keck Foundation, and the David and Lucile Packard Foundation. This research is supported by the National Science Foundation under grant AST~0908253 and AST~1211129 as well as by funding from the office of the Dean of the College of Arts and Science at Georgia State University. This research has made use of the SIMBAD database, operated at CDS, Strasbourg, France. 
Thanks are also extended to the U.\ S.\ Naval Observatory for their continued support of the Double Star Program. We very much appreciate the hard work of Isabelle Tallon-Bosc for the support for this project. This research has made use of the Jean-Marie Mariotti Center \texttt{LITpro} service co-developped by CRAL, LAOG and FIZEAU \citep{LITPRO}.
\footnote{LITpro software available at http://www.jmmc.fr/litpro}

\bibliographystyle{plainnat}

\clearpage

\begin{deluxetable}{rcrrrrrrr}
\tablewidth{0pt}
\tabletypesize{\footnotesize}
\tablecaption{CHARA SFP Observations \label{tab_sfpobs}} 
\tablehead{
\colhead{\textbf{System}} &
\colhead{\textbf{Set}} &
\colhead{\textbf{MJD}} &
\colhead{\textbf{B(m)}} &
\colhead{\textbf{$\vec{\theta}$($^{\circ}$)}} &
\colhead{\textbf{$\vec{\rho}$(mas)}} &
\colhead{\textbf{$BY$}} &
\colhead{\textbf{$\theta$}} &
\colhead{\textbf{$\rho$(mas)}}
}
\startdata

$\omega$~And& 1& 54786.29163 & 329.24 &   9.76 &  9.58 &      &        &      \\
          &  & 54786.29637 & 329.34 &   8.66 & 12.03 &      &        &        \\
          &  & 54786.38633 & 329.91 & 347.51 & 22.61 &      &        &        \\
          &  & 54786.39170 & 328.73 & 346.32 & 22.76 & 2008.8772& 117.03 & 35.24 \\
\hline
          & 2& 55105.79028 & VEGA   &        &       & 2009.7500& 232.57 & 24.68 \\
\hline
          & 3& 55111.30572 & 321.05 &  30.97 & 23.42 &      &        &        \\
          &  & 55111.35179 & 326.94 &  20.93 & 21.19 &      &        &        \\   
          &  & 55111.36393 & 274.44 & 136.19 &  9.62 &      &        &        \\   
          &  & 55111.42817 & 303.61 &  51.56 & 27.38 & 2009.7670& 245.23 & 28.61 \\  
\hline
          & 4& 55115.27366 & 278.40 & 335.77 &$\lesssim$5&     &     &        \\
          &  & 55115.28937 & 320.12 &  31.97 & 25.14 &      &        &        \\
          &  & 55115.32751 & 325.76 &  23.82 & 20.66 &      &        &        \\
          &  & 55115.37560 & 311.93 &  63.05 & 29.19 &      &        &        \\
          &  & 55115.42699 & 300.57 &  43.31 & 26.79 &      &        &        \\
          &  & 55115.43530 & 250.00 & 119.11 & 18.47 & 2009.7780& 246.85 & 29.20 \\
\hline
          & 5& 55154.73407 & VEGA   &        &       & 2009.8844& 273.55 & 38.67 \\
\hline
          & 6& 55438.41255 & 305.46 &  82.90 & 29.03 &      &        &        \\
          &  & 55438.42418 & 278.20 & 146.86 & 28.46 &      &        &        \\
          &  & 55438.46260 & 275.30 & 137.64 & 30.17 &      &        &        \\
          &  & 55438.46697 & 313.52 &  69.24 & 23.83 &      &        &        \\
          &  & 55438.47285 & 328.00 & 197.39 &$\lesssim$5& 2010.6626& 290.61 & 33.43 \\
\hline
          & 7& 55454.00139 & VEGA   &        &       & 2010.7038& 323.97 & 21.08 \\
\hline
          & 8& 55454.87917 & VEGA   &        &       & 2010.7061& 326.7  & 20.40 \\
\hline
          & 9& 55455.86944 & VEGA   &        &       & 2010.7088& 329.03 & 19.88 \\
\hline
          &10& 55457.91458 & VEGA   &        &       & 2010.7145& 334.6  & 18.76 \\
\hline
          &11& 55516.32097 & 257.29 & 302.44 & 30.37 &      &        &        \\
          &  & 55516.32602 & 301.64 & 229.43 & 24.33 &      &        &        \\
          &  & 55516.33092 & 329.67 &   0.80 &$\lesssim$5& 2010.8758&  92.36 & 34.30 \\
\hline
          &12& 55775.47784 & 278.49 & 338.78 & 19.43 &      &        &        \\
          &  & 55775.48110 & 302.59 & 265.04 & 34.22 &      &        &        \\
          &  & 55775.48486 & 320.66 & 211.40 & 15.59 &      &        &        \\
          &  & 55775.51149 & 277.74 & 324.20 & 22.11 &      &        &        \\
          &  & 55775.51527 & 311.27 & 256.74 & 32.87 &      &        &        \\
          &  & 55775.51874 & 325.45 & 204.45 & 10.63 & 2011.5854&  95.58 & 34.68 \\
\hline
          &13& 55781.49346 & 154.89 & 244.53 & 30.45 &      &        &        \\
          &  & 55782.46689 & 278.48 & 330.58 & 22.70 &      &        &        \\
          &  & 55782.47149 & 305.21 & 263.10 & 35.57 &      &        &        \\
          &  & 55782.47329 & 322.23 & 209.56 & 10.47 & 2011.6038& 100.54 & 36.77 \\
\hline
          &14& 55796.47406 & 276.15 & 319.35 & 31.81 &      &        &        \\
          &  & 55796.47892 & 313.38 & 251.43 & 27.16 &      &        &        \\
          &  & 55796.51211 & 270.17 & 311.09 & 33.49 &      &        &        \\
          &  & 55796.52028 & 310.61 & 240.55 & 23.13 &      &        &        \\
          &  & 55796.52417 & 329.21 &  10.04 &  6.65 & 2011.6429& 110.97 & 35.79 \\
\hline
          &15& 55800.97014 & VEGA   &        &       & 2011.6535& 114.5  & 34.2 \\
\hline
          &16& 55804.96875 & VEGA   &        &       & 2011.6647& 118.6  & 35.4 \\
\hline
          &17& 55805.88541 & VEGA   &        &       & 2011.6672& 118.9  & 33.1 \\
\hline
          &18& 55808.46680 & 272.65 & 313.75 & 31.24 &      &        &        \\
          &  & 55808.47141 & 312.76 & 245.12 & 17.29 &      &        &        \\
          &  & 55808.50026 & 264.68 & 306.72 & 30.95 &      &        &        \\
          &  & 55808.50324 & 307.61 & 236.24 & 13.31 &      &        &        \\
          &  & 55808.50705 & 329.49 &   6.49 & 13.86 & 2011.6757& 121.93 & 31.60 \\
\hline
          &19& 55829.47895 & 250.59 & 299.36 & 22.27 &      &        &        \\
          &  & 55829.50961 & 233.92 & 293.38 & 18.36 &      &        &        \\
          &  & 55829.51409 & 287.85 &  35.10 &  6.51 &      &        &        \\
          &  & 55829.51668 & 329.25 & 350.38 & 21.15 &      &        &        \\
          &  & 55829.52144 & 226.52 & 291.16 & 20.43 &      &        &        \\
          &  & 55829.52630 & 283.85 &  30.66 & 10.55 &      &        &        \\
          &  & 55829.53052 & 328.91 & 347.48 & 22.42 & 2011.7333& 144.56 & 23.69 \\
\hline
          &20& 55843.49748 & 216.41 & 288.40 &$\lesssim$5&    &      &        \\
          &  & 55843.50346 & 279.11 &  24.35 & 19.48 &      &        &        \\
          &  & 55843.50806 & 328.28 & 343.87 & 19.45 &      &        &        \\
          &  & 55843.51536 & 276.47 &  20.22 & 17.64 &      &        &        \\
          &  & 55843.52185 & 327.40 & 340.47 & 19.57 & 2011.7716& 191.92 & 20.37 \\
\hline
          &21& 55867.29442 & 313.52 &  69.06 & 22.35 &      &        &        \\
          &  & 55867.30024 & 245.53 & 311.21 &$\lesssim$5&    &      &        \\
          &  & 55867.30339 & 300.65 &  19.59 & 21.09 & 2011.8367& 222.95 & 23.90 \\
\hline
          &22& 56116.44410 & 246.26 & 357.45 & 16.04 &      &        &        \\
          &  & 56116.45118 & 256.82 &  53.23 & 20.39 &      &        &        \\
          &  & 56116.47894 & 246.99 & 347.41 & 17.32 &      &        &        \\
          &  & 56116.48838 & 274.66 &  46.99 & 20.55 & 2012.5189& 210.97 & 21.51 \\
\hline
          &23& 56131.45171 & 273.39 &  99.73 & 20.98 &      &        &        \\
          &  & 56131.45494 & 305.70 &  41.85 & 24.49 &      &        &        \\   
          &  & 56131.49109 & 296.31 &  88.93 & 26.17 &      &        &        \\
          &  & 56131.49745 & 318.16 &  33.86 & 25.23 &      &        &        \\
          &  & 56132.45407 & 276.90 &  98.03 & 23.39 &      &        &        \\
          &  & 56132.45923 & 308.15 &  40.61 & 24.86 &      &        &        \\
          &  & 56132.50179 & 278.50 & 152.50 &$\lesssim$5&    &      &        \\
          &  & 56132.50395 & 302.85 &  84.85 & 25.19 &      &        &        \\
          &  & 56132.50918 & 321.02 &  31.00 & 24.37 & 2012.5615& 242.73 & 27.70 \\
\hline
          &24& 56141.49189 & 249.38 & 144.81 & 15.34 &      &        &        \\
          &  & 56141.49717 & 246.79 &  85.11 & 34.11 &      &        &        \\
          &  & 56141.50164 & 242.69 &  23.27 & 19.20 &      &        &        \\
          &  & 56141.52016 & 250.27 &  79.45 & 31.13 &      &        &        \\
          &  & 56141.52248 & 244.42 &  18.37 & 17.65 & 2012.5875& 259.14 & 33.47 \\
\hline
          &25& 56148.43688 & 278.27 & 157.68 & 10.26 &      &        &        \\
          &  & 56148.44204 & 294.83 &  89.75 & 34.02 &      &        &        \\
          &  & 56148.44580 & 316.78 &  35.03 & 24.53 &      &        &        \\
          &  & 56148.47143 & 306.04 &  82.42 & 35.55 &      &        &        \\ 
          &  & 56148.47361 & 322.59 &  29.09 & 23.85 & 2012.6065& 261.45 & 35.76 \\
\hline
          &26& 56186.39405 & 277.28 & 142.40 & 28.84 &      &        &        \\ 
          &  & 56186.39996 & 312.62 &  74.18 & 35.44 &      &        &        \\ 
          &  & 56186.40423 & 312.90 &  73.41 & 34.07 &      &        &        \\ 
          &  & 56186.40659 & 326.81 &  21.28 & 11.10 &      &        &        \\ 
          &  & 56186.43604 & 272.06 & 133.06 & 28.83 &      &        &        \\  
          &  & 56186.44046 & 312.36 &  64.02 & 32.18 &      &        &        \\ 
          &  & 56186.44287 & 328.82 &  13.18 &  8.42 & 2012.7104& 274.51 & 38.01 \\
\hline
          &27& 56195.21723 & 227.81 & 131.78 & 36.77 &      &        &        \\  
          &  & 56195.22068 & 275.90 &  51.78 & 21.07 &      &        &        \\  
          &  & 56195.28974 & 277.82 & 163.08 & 20.49 &      &        &        \\  
          &  & 56195.29970 & 287.67 &  93.39 & 35.59 &      &        &        \\  
          &  & 56195.30585 & 313.91 &  37.18 & 13.40 & 2012.7347& 288.68 & 37.08 \\
\hline
          &28& 56244.28144 & 328.68 & 194.00 & 13.40 &      &        &        \\ 
          &  & 56244.38883 & 223.77 & 110.39 &$\lesssim$5&    &      &        \\  
          &  & 56244.39403 & 282.43 & 208.79 & 13.02 &      &        &        \\ 
          &  & 56244.39830 & 328.77 & 166.56 & 12.29 & 2012.8691&  18.54 & 13.64 \\
\hline
          &29& 56258.20213 & 312.46 & 254.56 & 21.69 &      &        &        \\   
          &  & 56258.20854 & 276.38 & 139.89 &$\lesssim$5&    &      &        \\  
          &  & 56258.21204 & 326.93 & 200.95 & 13.11 &      &        &        \\ 
          &  & 56258.22187 & 313.53 & 249.55 & 20.71 &      &        &        \\   
          &  & 56258.22972 & 328.07 & 197.05 & 16.21 & 2012.9070&  55.74 & 20.42 \\
\hline
HD~178911 & 1& 54254.47802 & 176.18 & 329.48 & 67.38 &      &        &        \\
          &  & 54255.47226 & 106.82 & 269.34 & 49.17 & 2007.4219& 315.85 &   71.1 \\
\hline
          & 2& 54287.38767 & 247.62 &   6.94 & 36.30 &      &        &        \\
          &  & 54287.39665 & 247.54 &   4.92 & 37.76 &      &        &        \\
          &  & 54287.40027 & 247.52 &   4.11 & 37.40 &      &        &        \\
          &  & 54287.45357 & 247.67 & 352.06 & 49.17 &      &        &        \\
          &  & 54287.46043 & 247.75 & 350.54 & 50.61 &      &        &        \\
          &  & 54287.46836 & 247.84 & 348.79 & 51.32 &      &        &        \\
          &  & 54287.47382 & 247.92 & 347.60 & 50.03 &      &        &        \\
          &  & 54287.48275 & 248.00 & 345.67 & 55.63 &      &        &        \\
          &  & 54289.27782 & 246.83 &  28.29 & 16.02 &      &        &        \\
          &  & 54289.28499 & 247.23 &  27.00 & 15.72 &      &        &        \\
          &  & 54289.35189 & 247.97 &  13.61 & 27.00 &      &        &    \\
          &  & 54289.36386 & 247.83 &  11.01 & 31.55 & 2007.5123& 309.36 &   66.7 \\
\hline
          & 3& 54392.09675 & 276.09 & 311.99 & 29.46 &      &        &        \\
          &  & 54393.16536 & 253.17 & 300.86 & 33.25 &      &        &        \\
          &  & 54393.19627 & 234.44 & 296.94 & 35.04 & 2007.7994& 269.81 &   41.8 \\
\hline
          & 4& 54605.41302 & 273.83 & 153.45 & 21.50 &      &        &        \\
          &  & 54605.44835 & 277.38 & 144.95 & 29.15 & 2008.3809&  84.42 &   62.4 \\
\hline
          & 5& 54692.26049 & 329.96 &  15.39 & 46.61 &      &        &        \\
          &  & 54692.30746 & 328.76 &   4.52 & 35.94 &      &        &        \\
          &  & 54692.31519 & 266.47 & 125.32 & 37.51 &      &        &        \\
          &  & 54692.37147 & 329.33 & 349.38 & 18.42 & 2008.6186&  65.69 &   73.7 \\
\hline
          & 6& 55115.12027 & 306.25 &  62.87 & 65.49 &      &        &        \\
          &  & 55115.17057 & 261.94 & 203.52 & 33.96 & 2009.7759&  13.82 &  100.2 \\
\hline
          & 7& 55346.42248 & 330.60 &  25.43 & 71.25 &      &        &        \\
          &  & 55346.49629 & 302.75 &  60.40 & 28.54 &      &        &        \\
          &  & 55347.50372 & 398.76 &  57.86 & 31.77 &      &        &        \\
          &  & 55348.33696 & 320.94 &  39.30 & 57.82 & 2010.4113& 349.10&   88.5 \\
\hline
          & 8& 55439.16025 & 310.57 & 258.16 & 22.89 &      &        &        \\
          &  & 55439.16753 & 330.58 &  25.70 & 48.07 &      &        &        \\
          &  & 55440.13761 & 274.90 & 331.13 & 84.67 &      &        &        \\
          &  & 55440.14199 & 306.16 & 261.22 & 28.09 &      &        &        \\
          &  & 55440.14627 & 329.91 &  29.27 & 45.60 & 2010.6648& 336.50&   83.4 \\
\hline
          & 9& 55683.42156 & 269.16 & 344.42 & 12.75 &      &        &        \\
          &  & 55683.42480 & 278.59 &  91.21 & 34.96 &      &        &        \\
          &  & 55683.43068 & 323.54 &  37.65 & 16.14 &      &        &        \\
          &  & 55683.46464 & 274.02 & 153.04 & 19.76 &      &        &        \\
          &  & 55683.46827 & 303.69 &  82.51 & 37.97 &      &        &        \\
          &  & 55683.47289 & 329.42 &  30.61 & 18.74 & 2011.3334& 272.75 &   39.0 \\
\hline
          &10& 55750.31281 & 277.25 & 145.35 & $\lesssim$5 &     &   &     \\
          &  & 55750.31537 & 312.00 & 256.64 & 20.51 &     &        &        \\
          &  & 55750.31904 & 330.62 & 204.94 & 27.56 & 2011.5164& 223.15 &   26.7 \\
\hline
          &11& 55775.23804 & 276.69 & 146.90 & 23.91 &     &        &        \\
          &  & 55775.24174 & 310.84 & 257.91 & $\lesssim$5 &     &  &        \\
          &  & 55775.24487 & 330.52 & 206.16 & 21.37 & 2011.5847& 168.89 &   26.2 \\
\hline
          &12& 55781.16531 & 176.82 & 172.22 & 32.23 &     &        &        \\
          &  & 55781.16903 & 145.91 & 258.19 & $\lesssim$5 &     &   &        \\
          &  & 55781.17239 & 245.07 & 211.91 & 16.83 &     &        &        \\
          &  & 55781.29349 & 176.45 & 150.41 & 23.91 &     &        &        \\
          &  & 55781.30175 & 247.59 & 186.23 & 26.18 & 2011.6010& 167.22 &   27.8 \\
\hline
          &13& 55872.30055 & 274.54 & 130.32 & 21.41 &     &        &        \\
          &  & 55872.30607 & 301.97 & 239.88 & $\lesssim$5 &   &    &        \\
          &  & 55872.30953 & 329.91 & 187.74 & 25.97 & 2011.6040& 163.22 &   26.9 \\
\hline
          &14& 55796.13238 & 271.32 & 158.97 & 25.93 &    &        &        \\
          &  & 55796.13608 & 291.49 &  87.30 & 15.98 &    &        &        \\
          &  & 55796.13942 & 326.65 &  34.91 & 11.75 &    &        &        \\
          &  & 55796.17761 & 276.43 & 147.55 & 27.34 &    &        &        \\
          &  & 55796.18115 & 310.21 &  78.47 & 11.61 &    &        &        \\
          &  & 55796.18441 & 330.44 &  26.81 & 15.77 & 2011.6420& 146.60&   28.6 \\
\hline
          &15& 55829.21391 & 262.59 & 123.76 & 38.91 &     &        &        \\
          &  & 55829.21798 & 283.22 &  48.92 & 18.39 &     &        &        \\
          &  & 55829.22325 & 328.64 & 178.01 & 19.64 & 2011.7324& 114.23 &   40.8 \\
\hline
          &16& 55843.12053 & 276.64 & 132.75 & 40.74 &     &        &        \\
          &  & 55843.12472 & 306.93 &  63.39 & 29.01 &     &        &        \\
          &  & 55843.21415 & 243.02 & 118.53 & 39.39 &     &        &        \\
          &  & 55843.21797 & 262.56 &  36.32 & 12.03 &     &        &        \\
          &  & 55843.22321 & 329.37 & 169.02 & 22.97 & 2011.7707& 110.72 &   42.2 \\
\hline
          &17& 56053.43557 & 148.95 &  76.07 & 78.33 &     &        &        \\
          &  & 56053.43978 & 176.95 & 352.70 & 42.81 &     &        &        \\
          &  & 56053.46060 & 247.45 &  26.14 & 73.31 &     &        &        \\
          &  & 56053.46509 & 154.22 &  70.67 & 80.67 &     &        &        \\
          &  & 56053.46718 & 177.24 & 346.38 & 34.27 & 2012.3460&  54.23 &  82.81 \\

\hline

$\xi$~Cep & 1& 53543.46805 & 264.15 & 145.30 & 43.01 &     &        &        \\
          &  & 53543.47206 & 263.59 & 144.16 & 46.18 &     &        &        \\   
          &  & 53543.47332 & 263.41 & 143.81 & 50.25 &     &        &        \\
          &  & 53545.47370 & 307.21 &  81.13 & 28.31 &     &        &        \\
          &  & 53545.48238 & 308.57 &  78.34 & 27.74 &     &        &        \\
          &  & 53545.49434 & 310.16 &  74.51 & 24.02 & 2005.4770& 134.74 & 47.05 \\
\hline
          & 2& 53605.26045 & 296.20 &  97.24 & 18.78 &     &        &        \\
          &  & 53605.26625 & 297.77 &  95.32 & 20.41 &     &        &        \\
          &  & 53605.27839 & 300.82 &  91.32 & 15.62 &     &        &        \\
          &  & 53605.28436 & 302.20 &  89.38 & 17.07 &     &        &        \\
          &  & 53605.29017 & 203.47 &  87.49 & 15.65 &     &        &        \\
          &  & 53606.26080 & 297.04 &  96.22 & 20.02 &     &        &        \\
          &  & 53606.27070 & 299.61 &  92.96 & 19.47 &     &        &        \\
          &  & 53606.28720 & 303.42 &  87.57 & 16.13 &     &        &        \\
          &  & 53606.29554 & 305.12 &  84.87 & 15.55 &     &        &        \\
          &  & 53606.30309 & 306.51 &  82.44 & 15.38 &     &        &        \\
          &  & 53607.21439 & 283.67 & 111.12 & 26.94 &     &        &        \\
          &  & 53607.22045 & 285.68 & 109.00 & 24.27 &     &        &        \\
          &  & 53607.26000 & 297.56 &  95.58 & 18.34 &     &        &        \\
          &  & 53607.26623 & 299.17 &  93.52 & 18.24 &     &        &        \\
          &  & 53607.27263 & 300.75 &  91.42 & 16.33 & 2005.6462& 149.64 & 32.98 \\
\hline
          & 3& 54956.46830 & 237.76 &  65.66 & 48.00 &     &        &        \\
          &  & 54956.47822 & 268.73 & 126.87 & 46.15 &     &        &        \\
          &  & 54956.48590 & 271.36 & 177.59 &$\lesssim$5& 2009.3430&  90.76 & 54.77 \\
\hline
          & 4& 55054.33274 & 215.59 &  26.97 & 17.70 &     &        &        \\
          &  & 55054.34009 & 216.80 &  24.89 & 16.13 &     &        &        \\
          &  & 55054.40038 & 222.38 & 189.93 &$\lesssim$5&     &        &        \\
          &  & 55055.28509 & 207.29 &  37.79 & 26.73 &     &        &        \\
          &  & 55055.28923 & 208.12 &  36.98 & 20.92 &     &        &        \\
          &  & 55055.29340 & 209.01 &  35.97 & 22.73 &     &        &        \\
          &  & 55055.37557 & 220.88 &  15.56 &$\lesssim$5&     &        &    \\
          &  & 55055.37994 & 221.23 &  14.46 &$\lesssim$5& 2009.6126& 101.87 & 57.39 \\
\hline
          & 5& 55111.34328 & 301.45 & 169.10 & 37.03 &     &        &        \\
          &  & 55111.38765 & 308.39 & 194.63 & 12.62 &     &        &        \\
          &  & 55111.42014 & 307.59 & 182.87 & 23.81 &     &        &        \\
          &  & 55115.25704 & 243.85 & 119.36 & 58.03 &     &        &        \\
          &  & 55115.28199 & 302.85 & 182.07 & 26.14 &     &        &        \\
          &  & 55115.36752 & 308.77 & 197.85 &  9.83 & 2009.7725& 117.67 & 58.40 \\
\hline
          & 6& 55438.36362 & 246.65 & 301.95 & 17.27 &     &        &        \\
          &  & 55438.36833 & 313.51 & 236.62 & 53.89 &     &        &        \\
          &  & 55438.37562 & 302.17 & 187.78 & 36.17 &     &        &        \\
          &  & 55438.43415 & 220.71 & 281.79 & 32.75 &     &        &        \\
          &  & 55438.43678 & 311.23 & 213.59 & 49.53 & 2010.6624& 231.84 & 52.25 \\
\hline
          & 7& 55445.15468 & 159.94 & 189.88 & 33.80 &     &        &        \\
          &  & 55445.15817 & 190.54 & 232.51 & 50.10 &     &        &        \\
          &  & 55445.16096 & 133.46 & 287.40 & 27.15 &     &        &        \\  
          &  & 55445.21604 & 206.99 & 218.23 & 49.73 & 2010.6810& 232.73 & 49.38 \\
\hline
          & 8& 55808.43387 & 216.69 &  97.81 & 64.91 &     &        &        \\
          &  & 55808.43643 & 310.52 &  29.34 & 25.49 &     &        &        \\
          &  & 55808.44095 & 301.06 & 167.75 & 20.26 & 2011.6756&  96.07 & 64.85 \\
\hline
          & 9& 55829.35906 & 222.01 & 102.67 & 66.09 &     &        &        \\
          &  & 55829.36364 & 248.44 &  38.73 & 30.14 &     &        &        \\
          &  & 55829.36755 & 245.04 & 164.84 & 26.88 &     &        &        \\
          &  & 55829.39391 & 205.69 &  92.04 & 64.86 &     &        &        \\
          &  & 55829.39839 & 246.21 &  26.67 & 17.60 &     &        &        \\
          &  & 55829.40049 & 240.86 & 156.47 & 36.77 & 2011.7329& 100.50 & 65.30 \\
\hline
          &10& 55843.33283 & 217.05 &  99.36 & 64.08 &     &        &        \\
          &  & 55843.33683 & 247.73 &  34.96 & 24.74 &     &        &        \\
          &  & 55843.34008 & 243.83 & 161.97 & 32.45 & 2011.7711& 102.31 & 64.18 \\
\hline
          &11& 55892.08110 & 256.78 & 133.23 & 58.82 &     &        &        \\
          &  & 55892.08577 & 311.75 &  69.51 & 51.06 &     &        &        \\
          &  & 55892.08983 & 298.41 &  18.97 &$\lesssim$5& 2011.9046& 108.18 & 65.13 \\
\hline
          &12& 56082.47286 & 268.72 & 157.91 & 36.61 &     &        &        \\
          &  & 56082.47817 & 239.11 & 102.02 & 18.91 &     &        &        \\
          &  & 56082.48203 & 234.60 & 211.85 & 22.50 &     &        &        \\
          &  & 56082.48501 & 267.71 & 154.38 & 34.86 &     &        &        \\
          &  & 56082.48968 & 241.40 &  98.10 & 18.01 &     &        &        \\
          &  & 56082.49370 & 236.95 & 209.07 & 25.80 & 2012.4259& 160.69 & 36.52 \\
\hline
          &13& 56195.17156 & 268.15 & 155.83 & 21.53 &     &        &        \\
          &  & 56195.17708 & 297.96 & 275.08 & 22.59 &     &        &        \\
          &  & 56195.18340 & 282.85 & 218.48 & 42.76 &     &        &        \\
          &  & 56195.26650 & 253.28 & 128.91 &$\lesssim$5&     &        &        \\
          &  & 56195.27046 & 312.76 & 244.77 & 38.53 &     &        &        \\
          &  & 56195.27504 & 300.08 & 195.12 & 38.15 & 2012.7345& 217.45 & 42.63 \\
\hline
          &14& 56245.21508 & 224.56 & 284.41 & 30.82 &     &        &        \\
          &  & 56245.22044 & 311.68 & 216.35 & 48.13 &     &        &        \\
          &  & 56245.22627 & 302.20 & 172.42 & 24.72 & 2012.8714& 232.62 & 50.00 
\\     
\hline 
\enddata
\tablecomments{Observation log for $\omega$~Andromeda, HD~178911, and $\xi$~Cephei on the CHARA Array from 2005 to 2012.  Each set of vector observations (along with the projected baseline length and epoch of observation) in columns $3-6$ were combined to create the true location of the secondary and average time of all the data points defined in the last three columns. Errors for all measurements in the final column are $\approx$ 1~mas. }
\end{deluxetable}

\clearpage

\begin{deluxetable}{lc}
\tablewidth{0pt}
\tabletypesize{\footnotesize}
\tablecaption{$\omega$~Andromeda Orbital Elements and Calculated Values. \label{tab_8799}} 
\tablehead{
\multicolumn{1}{l}{\textbf{Elements}} &
\multicolumn{1}{c}{\textbf{This Paper}}
}
\startdata
P (days)                 &  254.9003  $\pm$ 0.1960    \\
\ \ \ (yr)               &    0.69789 $\pm$ 0.00054  \\
T$_{\rm 0}$ (MJD)        &54214.835   $\pm$ 3.187    \\
\ \ \ \ \ (BY)           & 2007.3110  $\pm$ 0.0087    \\
a($''$)                    &    0.038   $\pm$ 0.001    \\
e                        &    0.142   $\pm$ 0.012    \\
i ($^{\circ}$)           &   62.49    $\pm$ 2.10    \\
$\omega$ ($^{\circ}$)    &  278.87    $\pm$ 2.01    \\
$\Omega$ ($^{\circ}$)    &  115.94    $\pm$ 4.38    \\
\hline
K$_{\rm 1}$ (km/s)       &   17.54    $\pm$ 0.30    \\
K$_{\rm 2}$ (km/s)       &   19.62    $\pm$ 0.30    \\
$\gamma_{\rm 0}$ (km/s)  &   14.83    $\pm$ 0.17    \\
\hline
$\chi_{\nu}^2$(RV)&  106.53 \\  
$\chi_{\nu}^2$(VIS)&  15.59 \\
$\chi_{\nu}^2$(Combined)&  84.29 \\
\hline
$M_{\rm P}$ ($M_{\odot}$)&    0.993   $\pm$ 0.056    \\
$M_{\rm S}$ ($M_{\odot}$)&    0.888   $\pm$ 0.058    \\
$\pi_{\rm orb}$ ($''$)   &    0.03912 $\pm$ 0.00197    \\
$\pi_{\rm Hip}$ ($''$)   &    0.03494 $\pm$ 0.0031    \\
\hline
\enddata
\end{deluxetable}

\clearpage

\begin{deluxetable}{lcc}
\tablewidth{0pt}
\tabletypesize{\footnotesize}
\tablecaption{HD~178911 Orbital Elements and Calculated Values. \label{tab_7272}} 
\tablehead{
\multicolumn{1}{l}{\textbf{Elements}} &
\multicolumn{1}{c}{\textbf{Tokovinin (2000)}} &
\multicolumn{1}{c}{\textbf{This Paper}}
}
\startdata
P (days)                 &   1296.3 $\pm$ 1.1 \phn &  1296.984 $\pm$ 0.355 \phn\phn\phn   \\
\ \ \ (yr)               &   \phn\phn 3.55 $\pm$ 0.003 & 3.55102 $\pm$ 0.00097\phn \\
T$_{\rm 0}$ (MJD)        &   50572.2 $\pm$ 1.5\phn\phn\phn & 50574.953 $\pm$ 1.302\phn\phn\phn \\
\ \ \ \ \ (BY)           & \phn\phn 1997.337 $\pm$ 0.00411\phn\phn & 1997.34538 $\pm$ 0.00356\phn\phn\phn\phn\\
a($''$)                    &    \phn 0.0735 $\pm$ 0.0026  &  0.074 $\pm$ 0.002    \\
e                        &   \phn  0.589 $\pm$ 0.004  &  \phn 0.597 $\pm$ 0.003 \phn \\
i ($^{\circ}$)           &   150.1 $\pm$ 3.7 \phn   &   147.29 $\pm$ 0.99\phn\phn      \\
$\omega$ ($^{\circ}$)    &   262.5 $\pm$ 0.8 \phn    &   83.88 $\pm$ 0.87\phn\phn       \\
$\Omega$ ($^{\circ}$)    &   276.7 $\pm$ 1.5 \phn   &   276.91 $\pm$ 1.45\phn      \\
\hline
K$_{\rm 1}$ (km/s)       & 6.57 $\pm$ 0.04 &   6.47 $\pm$ 0.09   \\
K$_{\rm 2}$ (km/s)       & 8.53 $\pm$ 0.17 &   8.33 $\pm$ 0.18   \\
$\gamma_{\rm 0}$ (km/s)  & -41.01 $\pm$ 0.03 \phn & -41.04 $\pm$ 0.06 \phn  \\
\hline
$\chi_{\nu}^2$(RV)& & 0.685 \\  
$\chi_{\nu}^2$(VIS)& & 2.187 \\
$\chi_{\nu}^2$(Combined)& & 0.997 \\
\hline
$M_{\rm P}$ ($M_{\odot}$)& 1.07 $\pm$ 0.37 & 0.802 $\pm$ 0.055   \\
$M_{\rm S}$ ($M_{\odot}$)& 0.84 $\pm$ 0.29 & 0.622 $\pm$ 0.053   \\
$\pi_{\rm orb}$ ($''$)   & 0.025 $\pm$ 0.008 & 0.02826 $\pm$ 0.00170\\
\hline
\enddata
\end{deluxetable}

\clearpage

\begin{deluxetable}{lcc}
\tablewidth{0pt}
\tabletypesize{\footnotesize}
\tablecaption{$\xi$~Cephei Orbital Elements and Calculated Values \label{tab_209790}}
\tablehead{
\multicolumn{1}{c}{\textbf{Elements}} &
\multicolumn{1}{c}{\textbf{Pourbaix (2000)}} &
\multicolumn{1}{c}{\textbf{This Paper}} 
}
\startdata
P (days)      &  818.51 $\pm$ 0.98 \phn\phn &  819.9402 $\pm$ 0.6082  \\
\ \ \ (yr)  &  \phn\phn 2.241 $\pm$ 0.0027 \phn &  2.24492 $\pm$ 0.00167   \\
T$_{\rm 0}$ (MJD)& 40949.584 $\pm$ 3.36 \phn\phn\phn\phn\phn  &  40949.144 $\pm$ 3.973 \phn\phn \\
\ \ \ \ \ (BY)& \phn\phn 1970.992 $\pm$ 0.0092\phn\phn\phn\phn & 1970.9908 $\pm$ 0.0105 \phn  \\
a($''$)         & 0.072 $\pm$ 0.0017  &  0.074 $\pm$ 0.004    \\
e             & 0.50 $\pm$ 0.021  &  0.481 $\pm$ 0.024      \\
i ($^{\circ}$)& 68 $\pm$ 1.4 &  70.96 $\pm$ 1.72     \\
$\omega$ ($^{\circ}$)& 273 $\pm$ 1.1 \phn   &  272.98 $\pm$ 1.95    \\
$\Omega$ ($^{\circ}$)& 85 $\pm$ 1.9   &  89.64 $\pm$ 3.51     \\
\hline
K$_{\rm 1}$ (km/s)       &   7.16 $\pm$ 0.56 &   7.81 $\pm$ 0.50  \\
K$_{\rm 2}$ (km/s)       &  19.82 $\pm$ 0.55 &  19.98 $\pm$ 0.83  \\
$\gamma_{\rm 0}$ (km/s)  & -10.74 $\pm$ 0.34 & -10.59 $\pm$ 0.33  \\
\hline
$\chi_{\nu}^2$(RV)& & 204.65 \\  
$\chi_{\nu}^2$(VIS)& & 45.01 \\
$\chi_{\nu}^2$(Combined)& & 150.55 \\
\hline
$M_{\rm P}$ ($M_{\odot}$)&  1.00 $\pm$ 0.13 & 1.045 $\pm$ 0.032\\
$M_{\rm S}$ ($M_{\odot}$)&  0.36 $\pm$ 0.05 & 0.409 $\pm$ 0.066\\
$\pi_{\rm orb}$ ($''$)   &  0.038 $\pm$ 0.0021\phn\phn & 0.03811 $\pm$ 0.00282\\

\hline
\enddata
\end{deluxetable}

\clearpage

\clearpage

\clearpage 

\clearpage

\clearpage

\clearpage

\end{document}